\documentclass[manuscript]{aastex61}
\usepackage{here}
\usepackage{booktabs}
\usepackage{amsmath,amssymb,amsfonts}
\usepackage{times}
\usepackage{lineno}
\shorttitle{Flare Statistics}
\shortauthors{Kawabata et al.}

\begin{document}
\title{Statistical relation between solar flares and coronal mass ejections with respect to sigmoidal structures in active regions}
\author{Yusuke Kawabata}
\affiliation{Department of Earth and Planetary Science, The University of Tokyo, 7-3-1 Hongo, Bunkyo-ku, Tokyo 113-0033, Japan}
\affiliation{Institute of Space and Astronautical Science, Japan Aerospace Exploration Agency, 3-1-1 Yoshinodai, Chuo, Sagamihara, Kanagawa 252-5210, Japan}

\author{Yusuke Iida}
\affiliation{Kwansei Gakuin University, Gakuen 2-1, Sanda, Hyogo 669-1337, Japan}

\author{Takafumi Doi}
\affiliation{Department of Earth and Planetary Science, The University of Tokyo, 7-3-1 Hongo, Bunkyo-ku, Tokyo 113-0033, Japan}
\affiliation{Institute of Space and Astronautical Science, Japan Aerospace Exploration Agency, 3-1-1 Yoshinodai, Chuo, Sagamihara, Kanagawa 252-5210, Japan}

\author{Sachiko Akiyama}
\affiliation{The Catholic University of America, Washington, DC 20064, USA}
\affiliation{NASA Goddard Space Flight Center, Greenbelt, MD 20771, USA}

\author{Seiji Yashiro}
\affiliation{The Catholic University of America, Washington, DC 20064, USA}
\affiliation{NASA Goddard Space Flight Center, Greenbelt, MD 20771, USA}

\author{Toshifumi Shimizu}
\affiliation{Institute of Space and Astronautical Science, Japan Aerospace Exploration Agency, 3-1-1 Yoshinodai, Chuo, Sagamihara, Kanagawa 252-5210, Japan}
\affiliation{Department of Earth and Planetary Science, The University of Tokyo, 7-3-1 Hongo, Bunkyo-ku, Tokyo 113-0033, Japan}
\email{kawabata.yusuke@ac.jaxa.jp}


\begin{abstract}
Statistical dependencies among features of coronal mass ejections (CMEs), solar flares, and sigmoidal structures in soft-X-ray images were investigated. 
We applied analysis methods to all the features in the same way in order to investigate the reproducibility  of the  correlations among them, which may be found from the combination of previous statistical studies. 
The samples of 211 M-class and X-class flares, which were observed between 2006 and 2015 by {\it Hinode}/X-ray telescope, {\it Solar and Heliospheric Observatory}/Large Angle and Spectrometric Coronagraph, and {\it GOES}, were examined statistically.   
Five kinds of analysis were performed: Occurrence rate analysis, linear-correlation analysis, association analysis, the Kolmogorov--Smirnov test, and Anderson-Darling test. The analyses show three main results. First, the sigmoidal structure and long duration events (LDEs) has stronger dependency on the CME occurrence than large X-ray class events in on-disk events. Second, for the limb events, the significant dependency exists between LDEs and CME occurrence, and between X-ray class and CME occurrence. Third, there existed 32.4\% of  on-disk flare events, which had sigmoidal structure and were not accompanied by CMEs. However, the occurrence probability of CMEs without sigmoidal structures is very small, 8.8 \%, in on-disk events.
While the first and second results are consistent with previous studies, we newly provided the difference between the on-disk events and limb events. The third result that non-sigmoidal regions produce less eruptive events is also different from previous results.
We suggest that sigmoidal structures in soft X-ray images will be a helpful feature for CME prediction 
regarding on-disk flare events.
\end{abstract}



\section{Introduction}
Coronal mass ejections (CMEs) are large amounts of plasma ejected from the solar corona into interplanetary space, thereby having a great influence on the space environment. 
The CMEs sometimes disturb the magnetosphere of the Earth, leading to catastrophic consequences for satellites and communication systems \citep{2006Natur.441..402C}. 
Other explosive events, known as solar flares, also occur in the solar atmosphere.
Solar flares are defined as an abrupt increase in electromagnetic radiation with a broad spectrum range.
 
 The relation between CMEs and solar flares is an important topic in solar physics. First, CMEs were considered to be initiated by large flares \citep{1976SoPh...50..153L}.
 However, \cite{2001ApJ...559..452Z} showed that CMEs are initiated before the onset of the associated flares. In their investigation, four CME events were analyzed with the Large Angle and Spectrometric Coronagraph \citep[LASCO;][]{1995SoPh..162..357B} and Extreme-ultraviolet Imaging Telescope \citep[EIT;][]{1995SoPh..162..291D} aboard the {\it Solar and Heliospheric Observatory} \citep[{\it SOHO};][]{1995SoPh..162....1D}. Scenarios in which CMEs are driven by flare-induced thermal pressure were rejected. 
 Although there seems to be a close relation between CMEs and solar flares, not all solar flares are accompanied by CMEs. 
 \cite{2012ApJ...755...44B} reported that 25$\%$ of CMEs were not associated with solar flares. 
 It is also known that solar flares without CMEs exist \citep{2006ApJ...650L.143Y}.
 Solar flares are roughly classified into two types: eruptive flares (flares with CMEs) and confined flares (flares without CMEs).
 Regarding eruptive flares, the eruption of a magnetic flux rope corresponds to a CME and the heated plasma caused by the magnetic reconnection is observed as a solar flare \citep[see the review by][]{2011LRSP....8....6S}.   
 Gas motion in the lower atmosphere, e.g., shear, rotation, and emergence, are thought to contribute to the storage of magnetic energy in the corona \citep{1982SoPh...79...59K, 1984SoPh...91..115H, 1996ApJ...462..547L} and to lead to solar flares and/or CMEs through certain initiation processes, e.g., breakout \citep{1999ApJ...510..485A}, tether cutting \citep{2001ApJ...552..833M}, preflare reconnection \citep{2012ApJ...760...31K, 2013ApJ...778...48B, 2017NatAs...1E..85W}, or magnetohydrodynamic instabilities \citep{1954RSPSA.223..348K, 2006PhRvL..96y5002K,2017ApJ...843..101I}.
 Although these models can explain specific flare--CME events, the question, which kind of solar flare (or coronal magnetic field configuration) is strongly related to the presence/absence of CMEs, is inconclusive in the statistical viewpoint.
 
Since coronal plasmas are heated to 10 MK--20 MK during solar flares, the characteristics of solar flares are often identified through X-ray observations \citep[e.g.][and references therein]{2017LRSP...14....2B}. The magnitude of a solar flare is usually defined by the peak flux of the soft-X-ray measured by  {\it Geostationary Operational Environmental Satellite} ({\it GOES}) and classified into several classes (A, B, C, M, and X). The light curves in soft X-rays are also characteristic features in solar flares, which can be divided into two phases: a rise and decay phase. The duration of solar flares ranges from several minutes to a few hours. Since the decay time corresponds to the plasma cooling time due to radiation and thermal conduction, the decay time is often used to diagnose the size of flaring loops \citep{2002ASPC..277..103R}. 
 
Several statistical observational studies have been done to investigate the properties of solar flares and/or CMEs. By focusing on flare occurrence, \cite{2015ApJ...798..135B} showed that a few parameters (total unsigned current helicity, total magnitude of Lorentz force, and total photospheric magnetic free energy density) are more relevant to the occurrence of solar flares than other parameters (mean photospheric magnetic free energy and mean current helicity). \cite{2015ChA&A..39..330Z}  indicated that the rising time of the soft-X-ray flux of a flare is approximately half of the decay time, and the rising time and decay time increase with variations in the peak flux of the soft X-ray. Regarding CMEs, \cite{2006ApJ...650L.143Y}  investigated the frequency distributions in the energy of solar flares and found that the power law indices of the frequency distributions for flares without CMEs are steeper than those for flares with CMEs. According to \cite{2017ApJ...834...56T}, if the area of flare ribbons normalized by the sunspot area is large, solar flares tend to be accompanied by CMEs. 

Another important observational characteristic of solar flares is the shape of the magnetic field lines in the corona. Since the plasmas in the corona are frozen to the magnetic field lines, the loop-like structure, which can be seen through soft X-ray or extreme ultraviolet (EUV) observations, reflects the structure of magnetic-field lines. One of the characteristic coronal structures in flare-productive active regions is a sigmoid---an S-shaped or inverse S-shaped bright structure visible in soft X-ray and EUV high temperature line images \citep{1996ApJ...464L.199R, 2007ApJ...669.1372L, 2014ApJ...789...93C}. A sigmoidal structure is thought to be formed by flux emergence \citep{2009ApJ...691.1276A} or flux cancellation \citep{1989ApJ...343..971V, 2012ApJ...759..105S}. Since sigmoids are bundles of twisted magnetic field lines containing electric currents and thus free energy, sigmoids are likely to be eruptive \citep{1999GeoRL..26..627C}.  
\cite{2014SoPh..289.3297S} reported that 64$\%$ of sigmoids produce flares and H${\rm \alpha}$ filaments are observed in 65$\%$ of sigmoids with 72 sigmoids analysis. Conventionally, sigmoids are identified by visual inspection, including the study of \cite{2014SoPh..289.3297S}. However, with increasing sample number, the identification of sigmoids requires more time and effort. In this study, we developed a new method for detecting the shape of bright X-ray regions and supporting the identification of sigmoids.

Some studies focused on comparison between CMEs and solar flares.
 \cite{2015SoPh..290..579D} aimed to investigate the relation among the geomagnetic $Dst$, the CME parameters, and solar flare parameters. Their results confirmed that the initial CME speeds, apparent width, position, and class of solar flares are related to the magnitude of geomagnetic storms. \cite{2015GeoRL..42.5702T} also studied the relation between the CME speeds and the magnetic parameters such as magnetic flux, magnetic twist, and free magnetic energy proxies. 
They showed that global alpha parameter \citep{2009ApJ...700..199T} determines the CME speeds upper limit.
 In detail, few studies focused on the statistical relation between the existence of sigmoids and the occurrence of CMEs, whereas \cite{1999GeoRL..26..627C} investigated the eruptivity of sigmoids by using only soft-X-ray data with {\it Yohkoh} \citep{1991SoPh..136...37T}. 
In this study, we focused on the relation between coronal parameters, e.g., X-ray class, duration of solar flares, and existence of sigmoids and CME parameters, e.g., existence, speed, and size.
We applied data obtained by the {\it Hinode} satellite \citep{2007SoPh..243....3K}, {\it SOHO}, and {\it GOES}.
All missions have been observing the Sun for more than ten years and enable statistical studies with enormous and uniform data quantities.
Although previous statistical studies investigated relations among CMEs, solar flares and sigmoidal structures, any of them did not investigate all of the relations among them in the same manner.  
Thus, the purpose of this investigation is to confirm the statistical relations among all of the features or parameters, by applying common analysis methods to the dataset.


This paper is organized as follows: The observations and data selection are described in Section 2, and the analysis is explained in Section 3. Section 4 presents the results, followed by  discussion in Section~5.   


\section{Data Selection and Observations}

The events were selected based on the flare observations.
We used the {\it Hinode} flare catalogue \citep{2012SoPh..279..317W}, which describes whether each solar flare was observed by the {\it Hinode} instruments. A set of 211 flare events were selected from a time period between December 2006 and June 2015 based on the following criteria: \\
(i) Larger than M-class flares (according to the soft-X-ray flux peak measured with {\it GOES}).  \\
(ii) There existed at least one map observed with the spectropolarimeter \citep[SP;][]{2013SoPh..283..579L}, which is one of two focal-plane instruments of the Solar Optical Telescope \citep[SOT;][]{2008SoPh..249..167T,2008SoPh..249..221S,2008SoPh..249..197S,2008SoPh..249..233I} aboard the {\it Hinode} satellite, within 6 h before and after the flare peak.  

The first criterion was set in order to focus on the relation among large flares, CMEs, and sigmoids.
The second criterion aims to investigate quantitatively the properties of photospheric magnetic field in relation to the occurrence of flares and CMEs  in our subsequent studies.  

To identify sigmoidal structures, we analyzed soft-X-ray images observed with the X-ray telescope \citep[XRT:][]{2007SoPh..243...63G, 2008SoPh..249..263K} aboard the {\it Hinode} satellite.
Recent observations show that sigmoids can be identified not only through soft X-ray but also through EUV observations \citep{2007ApJ...669.1372L}, which means that sigmoids have a wide range of temperature (1 MK--10 MK). However, the sigmoidal structure is most prominent and easily identified in soft-X rays \citep{2014SoPh..289.3297S}, which we use in this study. The XRT can observe coronal plasmas in a wide temperature range by using several filters. The majority of the chosen flare events were observed with the Be-thin, Ti-poly, and Al-thick.
Since the temperature sensitivity of Al-thick is extremely high, approximately 10 MK, the sigmoidal structure looks blurred. The Ti-poly filter is often saturated in the flaring region of the chosen events. Thus, images with Be-thin were used in our study. Each map consists of $384 \times 384$ pixels or $512 \times 512$ pixels  at a plate scale of $1\arcsec. 0286$ per pixel. The time cadence of the observations is different at each event (1--5 min). 

To derive the duration of flares, we used time series of the 1--8 \AA \ soft-X ray flux observed by {\it GOES}, which obtains data at 2 s or 3 s intervals ({\it GOES} 13--15 or {\it GOES} 11--12, respectively) and provides us with a continuous soft-X-ray flux during solar flares.


\section{Analysis}

\subsection{Property Definition of Events}
The focus of our analysis is to reveal the statistical relation among physical parameters of CMEs and solar flares simultaneously. We chose the CME presence/absence, speed, and size as the CMEs parameters, and X-ray class, duration, and sigmoid presence/absence as solar-flare parameters. Each characteristic is defined below.
\subsubsection{CMEs}
To derive the existence and physical parameters of CMEs, we used the {\it SOHO}/LASCO catalog\footnote{https://cdaw.gsfc.nasa.gov/CME\_list/} \citep{2004JGRA..109.7105Y, 2009EM&P..104..295G}.
In the {\it SOHO}/LASCO catalog, the existence of CMEs is manually identified from LASCO C2 and C3 data, which cover the outer corona for 2.5--6.0 solar radii (C2) and 3.7--30 solar radii (C3). The CME speed is defined by fitting a straight line (i.e., linear or first-order polynomial fit) to the height-time measurements, and the CME width is measured in the LASCO C2 after the width becomes stable.

\subsubsection{Duration of Flare Events}
We defined the flare rise time ($t_{\rm rise}$), decay time ($t_{\rm dec}$), and duration ($t_{\rm dur}$) as follows: 
\begin{eqnarray}
t_{\rm rise}&=&T_{\rm peak}-T_{\rm st}, \\
t_{\rm dec}&=& T_{\rm en}-T_{\rm peak}, \\
t_{\rm dur}&=&t_{\rm rise}+t_{\rm dec},
\end{eqnarray} 
where $T_{\rm st}$, $T_{\rm peak}$, and $T_{\rm en}$ are the flare start time, flare peak time, and flare end time, respectively.
Although $T_{\rm st}$, $T_{\rm peak}$, and $T_{\rm en}$ are provided by the {\it GOES} event list\footnote{http://www.lmsal.com/solarsoft/last\_events/}, we re-determined them based on the method of \cite{2012ApJ...754..112A}, which enabled us to distinguish more clearly between long duration events (LDEs) and impulsive flares. 
In the {\it GOES} event list, the flare end time is defined as time when the flux level decays to a point halfway between the maximum flux and pre-flare background level. 
However, this study defines the flare end time as time when the flux level reaches approximately the pre-flare background level.
The procedure is as follows:

1. {\it Data rebinning.} The original time resolution of $dt =$ 2 s or 3 s was rebinned to time steps of $\Delta t =$ 6 s. 

2. {\it Data smoothing.} The rebinned light curve was smoothed with a boxcar average, and the width of the smoothing window measured 60 s.

3. {\it Detection of flare peak time.} We defined the peak time, $T_{\rm peak}$, as the local maximum, where the X-ray flux, $F(T)$, satisfies $F (T_{\rm peak}-\Delta t)<F(T_{\rm peak})$, $F(T_{\rm peak}+\Delta t)<F(T_{\rm peak})$, and $F(T_{\rm peak}) > 10^{-5} {\rm Wm^{-2}}$.

4. {\it Detection of flare start time.} The definition of the flare start time ($T_{\rm st}$) is equal to that of the GOES start time, which is determined by the first minute of a sequence of 4 min of steep monotonic increase.

5. {\it Detection of flare end time.}  We defined the flare end time ($T_{\rm end}$) as the time when the soft-X-ray flux satisfies $F<3F_{\rm back}$, with $F_{\rm back}$ as the background flux, which is defined by the average X-ray flux 36 s before the flare start time.

\begin{figure}[H]
\centering
 \includegraphics[width=0.8\columnwidth,clip,bb=0 0 1023 768]{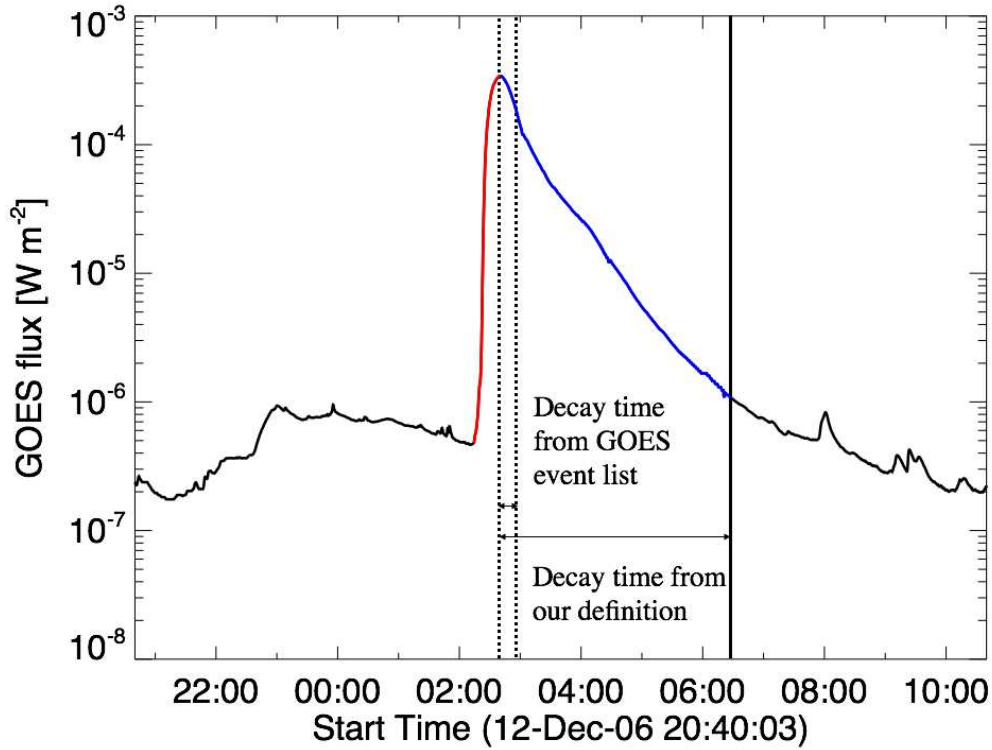}
 \caption{Soft-X-ray flux observed with {\it GOES} during period between 20:40 UT on 12 December 2006 and 10:40 UT on 13 December 2006. The red and blue lines show the rise time and decay time from the definitions, respectively. The peak time and end time from the {\it GOES} event list are represented by a vertical dotted line. The background flux in this event is $3\times 10^{-7} {\rm Wm^{-2}}$}.
 \label{duration}
\end{figure}

Figure \ref{duration} shows an example of {\it GOES} X-class flares and our detection of the rise time (red line), decay time (blue line), and duration (red line + blue line) described above. 
The peak time and end time from the {\it GOES} event list are represented by the vertical dotted line. 
The decay time of the {\it GOES} event list is quite short ($\sim$16 min), despite of visual inspections that conventionally declared this event as an LDE. 
In contrast, our method captured a long decay time ($\sim$4 h).

It should be noted that there are 20 events, which occurred at a decay time of other events. 
Further, five events lack data and were thus excluded from our duration analysis.

\subsubsection{Sigmoids}
\begin{figure}[H]
\centering
 \includegraphics[width=\columnwidth,clip]{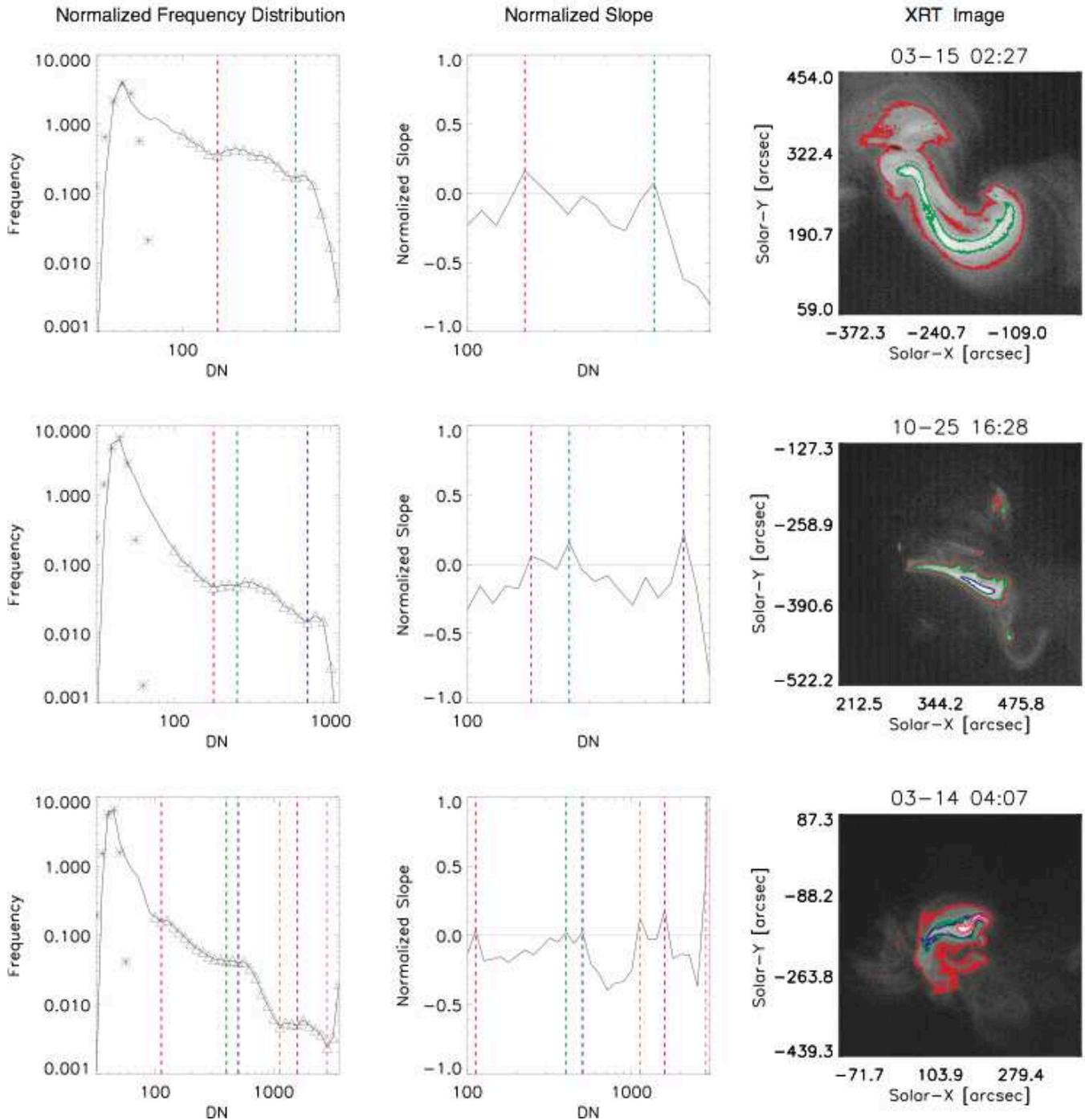}
 \caption{Examples of sigmoid structures in X-ray images. The left column shows the frequency distribution of the X-ray intensity in an image. The asterisks were fitted with a Gaussian function and indicate the photon noise. The vertical dashed lines correspond to signals presenting the enhancement begin. The middle column shows the derivative of the frequency distribution of the X-ray intensity. The horizontal line corresponds to zero and the vertical lines correspond to the lowest signals of each enhancement range. The right column presents the results of the bright-region detection. The background represents the X-ray intensity image and the contours show discriminated bright regions. The contour colors correspond to those in the left and middle columns.}
 \label{fig:sgm_ex}
\end{figure}

The sigmoids can be identified as S- shaped or inverse S-shaped structures with higher intensity than the surrounding structures in soft X-ray. We identified sigmoidal structures in soft-X-ray images as follows: 

First, we determined the frequency distribution of the X-ray intensity in each image to obtain the XRT noise level. The XRT data was normalized using the exposure time.
The left panels in Figure \ref{fig:sgm_ex} show examples of the frequency distribution at this step.
A peak exists in the range below 50 DN/s, which corresponds to photon noise.
We determined this value by fitting the distribution with a Gaussian function for each image.
The typical value of noise is 15 DN/s. Note that the system gain of the XRT is 57 ${\rm e^{-1}/DN}$.

Next, we looked for enhancement in the frequency distribution.
While the distribution typically decreases as a power law function in the range above the photon noise level, certain enhancement ranges are occasionally seen.
We extracted all enhancements in the range above 100 DN/s, which is marked with triangles in the left column of Figure \ref{fig:sgm_ex}, as ranges where the first derivative of the frequency distribution is larger than zero.
The middle column of Figure \ref{fig:sgm_ex} shows examples of this step.
The horizontal line corresponds to zero and the vertical dashed lines correspond to the lowest values of enhancement ranges detected in the analysis.
Afterwards, the bright enhancement region (candidate for sigmoid detection) was drawn in an X-ray intensity image using the threshold value obtained in the analysis above.
The right column of Figure \ref{fig:sgm_ex} presents an example of enhancement regions found with this method.
The colors of the contour lines correspond to those of the vertical dashed ones in the left and right columns.

At the final step of sigmoid recognition, we judged whether the bright regions are S- or J-shaped. Three authors, i.e., Y. Kawabata, Y. Iida, and T. Doi, separately carried out this step by our eyes for all bright regions of all flare events.
When two or three persons judge that the bright region has an S- or a J-shaped structure, the bright region is classified as a sigmoid.
The flare event is judged to be accompanied by a sigmoid structure when it exists at least in one image taken in 12 hours before the flare peak time. 

\subsection{ Association among Flare Parameters, Sigmoid Existence, and CME Occurrence}
We investigated the independency of four parameters  through contingency table and probabilities of intersection sets: X-ray class, duration, sigmoid (hereafter SGM) existence, and CME existence. We define large X-ray class (hereafter LXC) events as events in which the X-ray class is more than M2.3. Long duration events are defined as events in which the duration exceeds 1 h. Both thresholds were approximately set to the median value of the 211 and 186 events, respectively.

 First, we computed the occurrence probability of each event.
The events without Be-thin data were excluded from the sigmoid analysis.
Further, CME data was excluded when it was difficult to judge whether CMEs occurred or not.
An error was obtained as Poisson noise value in the case of many events, i.e., $\sqrt{N}$, with N as event number.

Next, the contingency table of the pairs among all four parameters, e.g. LXC, LDE, SGM , and CME, was created to investigate the association of each pair of parameters. We calculated the phi coefficient defined by             
\begin{equation}
\phi = \sqrt{\frac{\chi^2}{N}},
\end{equation}
where $\chi$ is Pearson's chi-square and N is the number of events.

 The association was also investigated from the different point of view. We focused on how large differences exist between the actual occurrence probabilities of the intersection sets and those assuming the independency of each parameter. We computed the occurrence probability of each intersection set, A $\cap$ B, from the individual occurrence probabilities by assuming the independency.
The occurrence probability of the intersection set can be obtained as a product due to the independent events in this case.
The error bars of the intersection events, $\sigma_{\rm ind}$, was calculated as Poisson noise with each occurrence probability. The occurrence probability of the actual intersection event was obtained from the observation result.
The error value, $\sigma_{\rm obs}$, was calculated as Poisson noise from the occurrence number of the actual intersection events. Finally, we compared the occurrence probability of the intersection events with independent events with that from the actual observation obtained in the analysis above, and calculated by how many multiples of the error value they are separated.

\subsection{Kolmogorov--Smirnov Test and Anderson-Darling test}
\label{ks_method}

\ In order to estimate how much sigmoid and CME occurrence depends on the four flare properties, i.e., (i) flare duration, (ii) rise time, (iii) decay time, and (iv) soft-X-ray flux, we conducted the Kolmogorov--Smirnov (KS) test and Anderson-Darling (AD) test for each condition. The KS and AD tests judge whether two distributions are similar or not. We derived the cumulative distribution functions (CDFs) of the four parameters (i)--(iv), 
\begin{equation}
F(x)=\int_{0}^{x} f(t)dt,
\end{equation}
where F and f are CDF and frequency distribution. The both of x and t are one from parameters (i) -- (iv).
The KS and AD tests were conducted for the two CDFs with and without CMEs (or sigmoids). 
The KS  and AD statistics, $D_{\rm KS} \ {\rm and}\  D_{AD}$ are calculated as, 
\begin{eqnarray}
D_{\rm KS}&=&{\rm sup}|F_1(x)-F_{2}(x)| \ \ \ , \\
D_{\rm AD}&=&\frac{mn}{m+n} \int \frac{F_1(x)-F_2(x)}{G(x)(1-G(x))}dG(x) \ \ \ ,
\end{eqnarray}
where sup is the supremum, $m$ and $n$ are the sample number of each distribution, and $G(x)=mF_{1}(x)-nF_{2}(x)$. 
The AD test gives more weight to the tails of the CDFs than the KS test.
 In the KS test, the two populations are judged to be different distributions in the confidence interval of $100(1-\alpha)$\% when the $D_{\rm KS}$ satisfies following relation,
 \begin{equation}
   D_{\rm KS} \geq \sqrt{\frac{n + m}{nm}}  \sqrt{-\frac{1}{2}\ln{\frac{\alpha}{2}}} \ \ \ .
  \label{kseq1}
 \end{equation}
 Note that the critical value is proportional to $\sqrt{(n + m)\diagup nm}$. The minimum value of $\alpha$ (e.g., $\alpha\sim0$) corresponds to the probability at which the two populations are in the different distribution function (i.e., $\sim$100$\%$ confidence), which means that the CMEs (sigmoids) occurrence have some dependency on the flare parameter. 
 In this study, we adopted $\alpha=0.01$ to judge whether the two populations are different or not.
 In the AD test, the critical value is also set to 99\% confidence as follows \citep{scholz1987k},
 
 \begin{equation}
 D_{\rm AD} \geq 3.752 \ \ \ .
 \end{equation}


\section{Results}

\subsection{General Description}

\begin{figure}[H]
\centering
 \includegraphics[width=\columnwidth,clip]{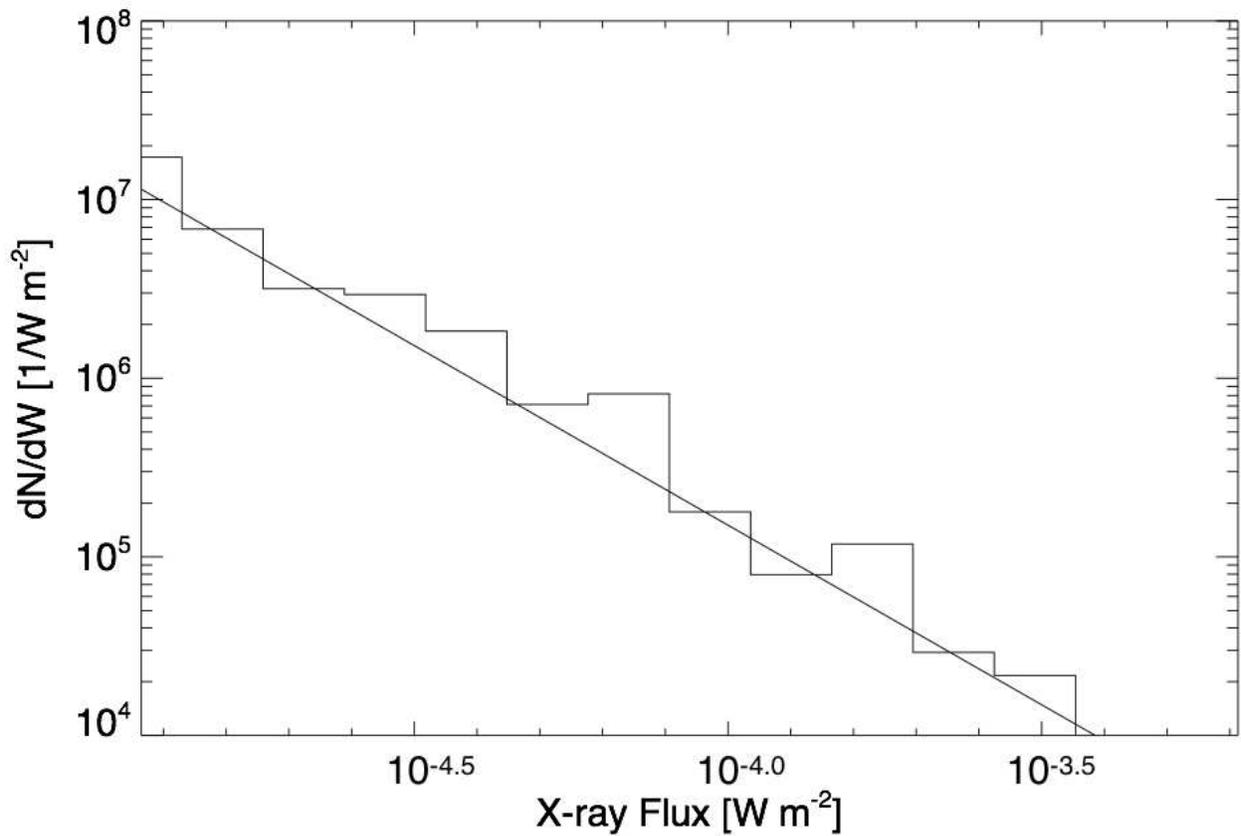}
 \caption{Frequency distribution of solar flares as function of soft-X-ray peak intensity from GOES flare list. The power law index of the distribution is $\sim -2.0$}
 \label{xray_histo}
\end{figure}

Figure \ref{xray_histo} presents the frequency distribution of solar flares as a function of the soft-X-ray peak intensity, $dN/dW$, which includes 195 M-class flares and 16 X-class flares. The median of the X-ray class is M2.3. The power law index of this distribution is $\sim -2.0$, which was derived by linearly fitting the frequency distribution. The value is consistent with the previous result ($\sim -1.9$) derived from the XRT observations \citep{2013ApJ...775...22S}.

\begin{figure}[H]
\centering
 \includegraphics[width=\columnwidth,clip]{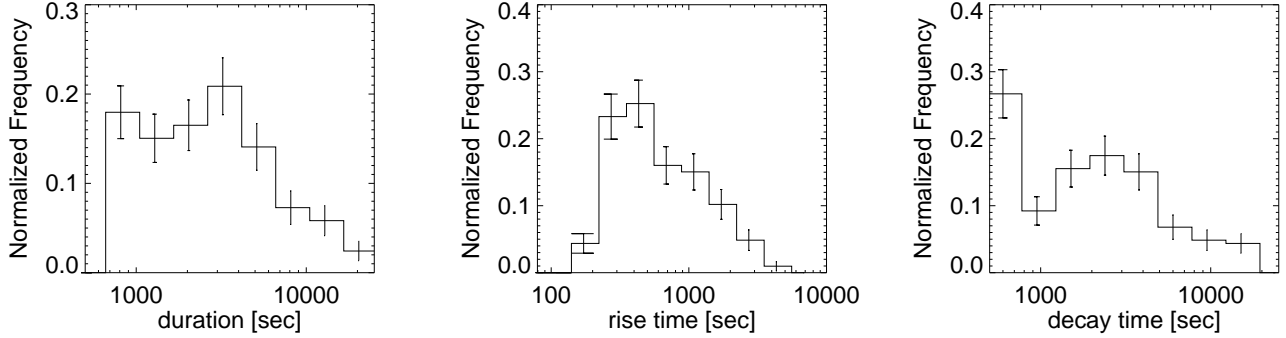}
 \caption{Histogram of duration, rise time, and decay time.}
 \label{duration_histo}
\end{figure}

Figure \ref{duration_histo} shows the histograms of the duration, rise time, and decay time. The respective medians are 3385 s, 676 s, and 2409 s. The duration and decay time show similar distributions. More events exist in the shorter time scale ($< 1000$ s) in the decay time histograms. Further, extremely long durations and decay time such as $\sim10000$ s exist. The rise time behaves differently from the duration and decay time. It has a peak at $\sim 400$ s.

We categorized the 211 events into two groups: on-disk and limb events. On-disk events are located within 500\arcsec \ from the disk center of the Sun and limb events describe the rest. Based on our definition, 63 on-disk events and 148 limb events occurred. The existence of sigmoids and CMEs for on-disk and limb events is summarized in Table \ref{n_event_sgmcme}. If no Be-thin images were taken in 12 h before the onset of flares, sigmoid judgement was impracticable. Furthermore, when it was difficult to identify which flare produced the CME, CME judgement was also impracticable.

\renewcommand{\arraystretch}{0.8}
\begin{table}[H]
\centering
\caption{Number of events at each observation position}
 \begin{tabular}{cccc} 
\toprule
 &  All  & On-disk  & Limb  \\
 &       (211 events)      &(63 events)          &(148 events)\\
\toprule
sigmoid & 58 & 21 & 37 \\
no sigmoid & 62 & 21 & 41 \\
undeterminable & 91 & 21 & 70 \\
\hline
CME & 57 & 15 & 42 \\
no CME & 120 & 36 & 84 \\
undeterminable & 34 & 12 & 22 \\
\bottomrule
\end{tabular}
\label{n_event_sgmcme}
\end{table}
\renewcommand{\arraystretch}{1}

\subsection{Occurrence Dependency Between Parameters}

In this section, we illustrate the dependency of sigmoid and CME occurrence on the duration, rise time, decay time, and X-ray flux of flares. The sigmoid (CME) occurrence is defined as the number ratio of events in which the sigmoid (CME) occurred. In the 211 events, 58 events in which sigmoids are observed and 62 events without sigmoids exist. Further, 57 events in which CME occurred and 120 events without CME exist (see Table \ref{n_event_sgmcme}). 

\begin{figure}[H]
\centering
 \includegraphics[clip,width=\columnwidth]{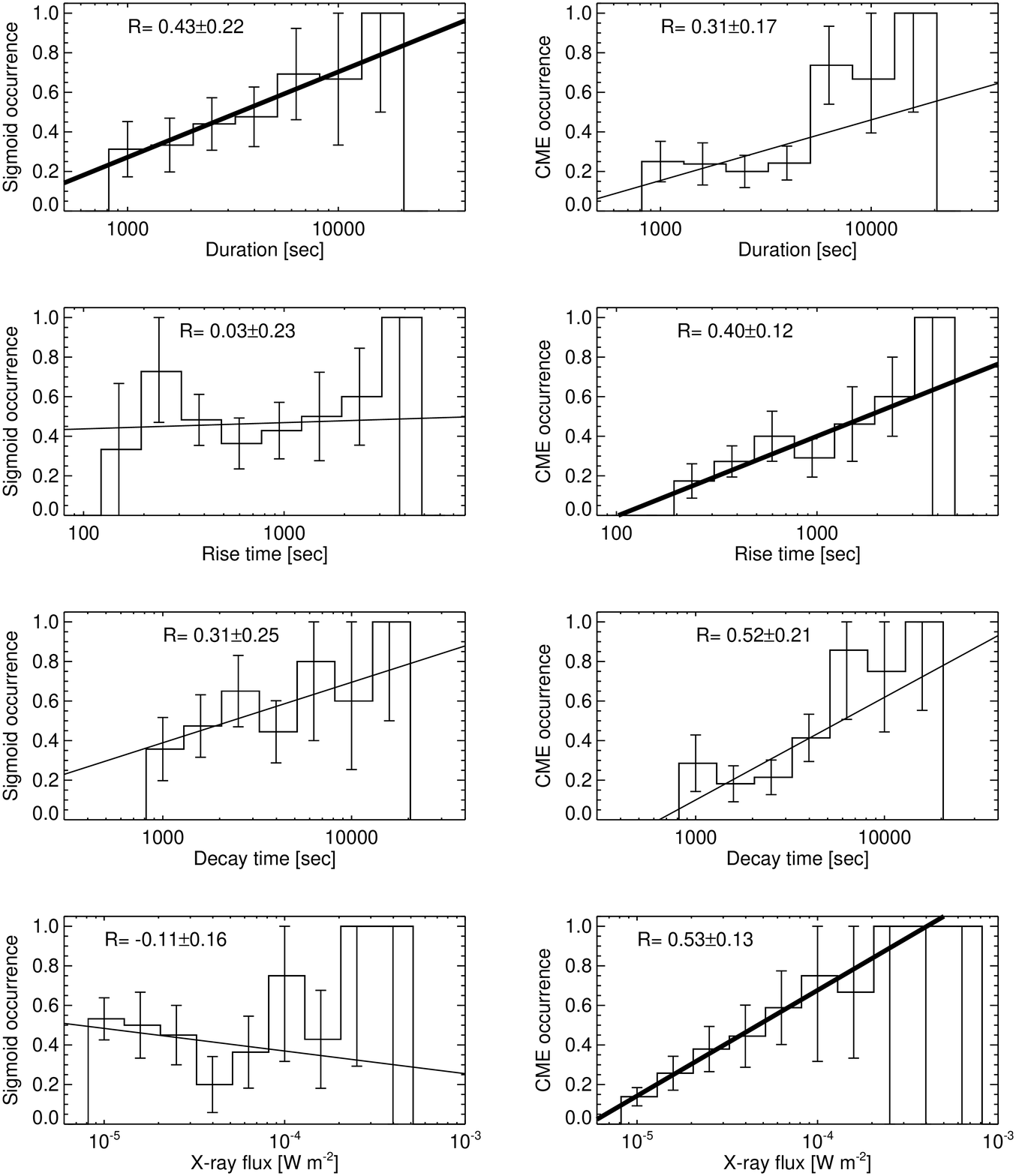}
 \caption{Sigmoid (left side) and CME (right side) occurrence as functions of flare duration, rise time, decay time, and soft-X-ray peak flux for all events in which recognizable sigmoids / CMEs occurred. The error bar length was estimated assuming Poisson noise. The inclined line represents the result of a linear fitting.}
 \label{histo_occurrence_211}
\end{figure}

\begin{figure}[H]
\centering
 \includegraphics[clip,width=\columnwidth]{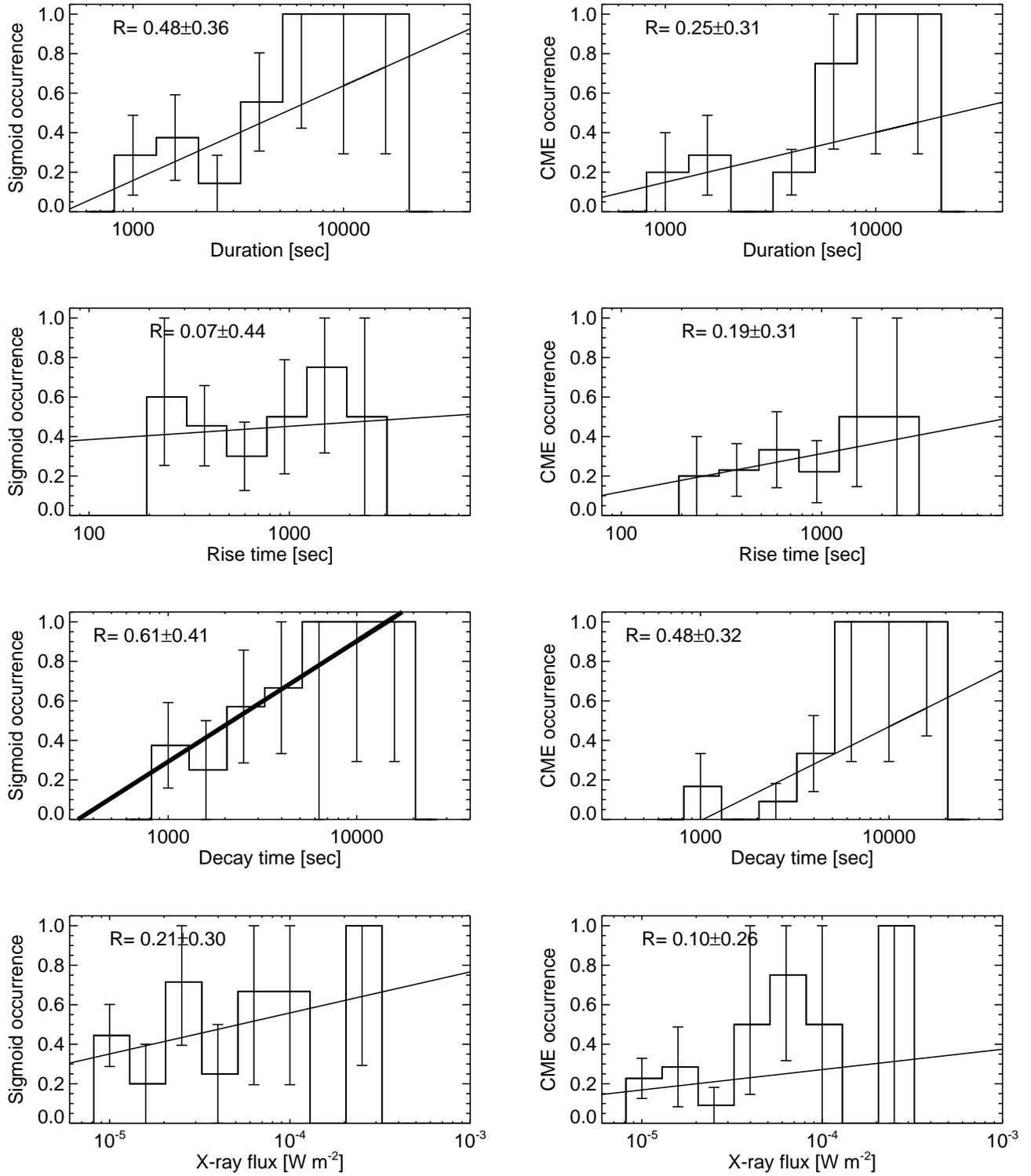}
 \caption{Similar to Figure \ref{histo_occurrence_211}, but covers on-disk events.}
 \label{histo_occurrence_center}
\end{figure}

\begin{figure}[H]
\centering
 \includegraphics[clip,width=\columnwidth]{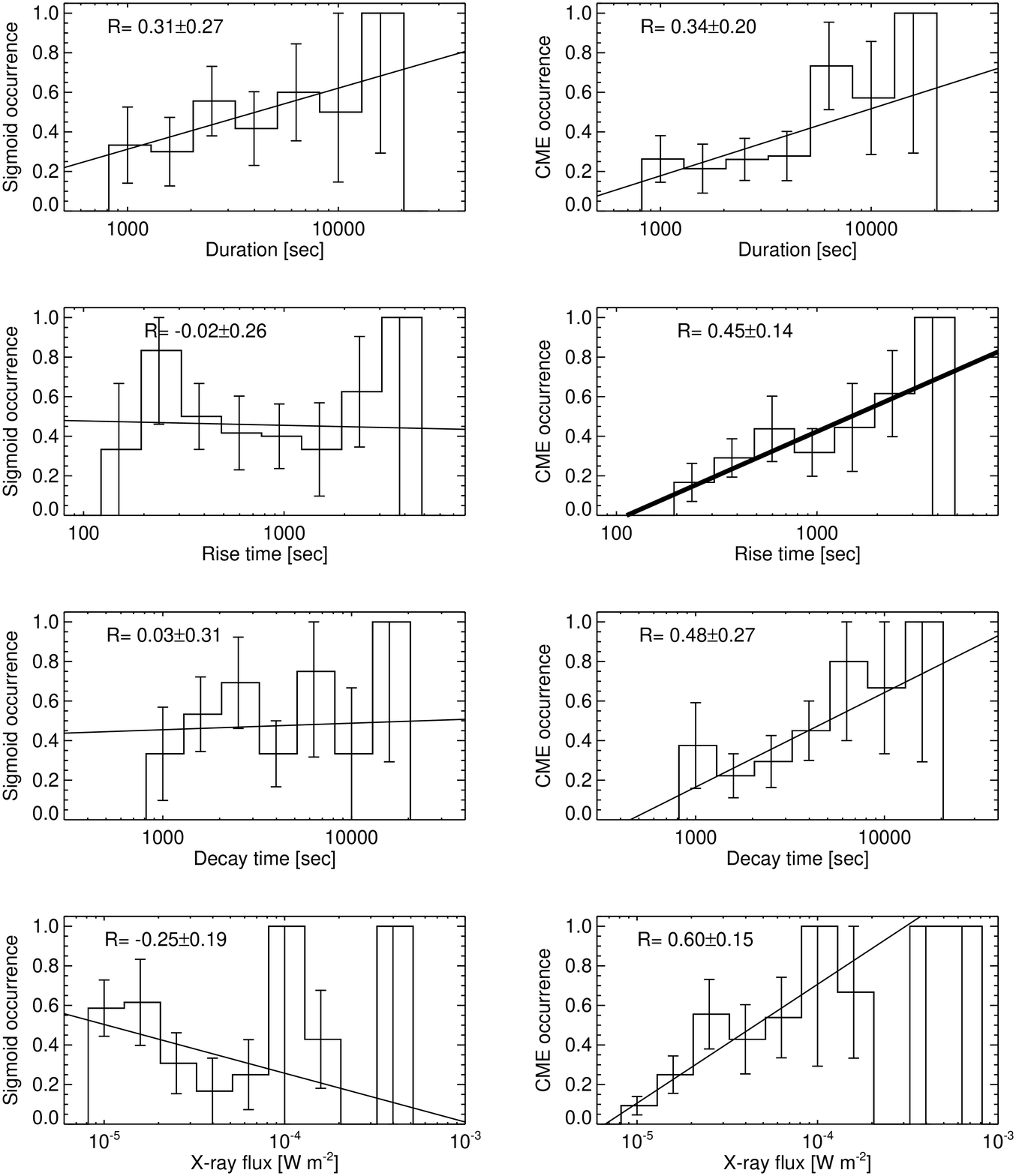}
 \caption{Similar to Figure \ref{histo_occurrence_211}, but covers limb events}
 \label{histo_occurrence_limb}
\end{figure}

In Table \ref{rho_tau}, we show the Spearman's rank correlation coefficient $\rho$, and Kendall's rank correlation coefficient $\tau$ of the occurrence dependencies, which are shown in Figures \ref{histo_occurrence_211}, \ref{histo_occurrence_center} and \ref{histo_occurrence_limb}. 
Bold values show the deviation from zero in 99\% confidence interval, which means that the relation of two parameters can be represented by monotonic function.
There are significant correlation between the duration and  sigmoid, rise time and CME, and X-ray flux and CME for all events.
Significant correlation also can be seen between decay time and sigmoid for the on-disk events, rise time and CME, and X-ray flux and CME for the limb events. 

\begin{table}[H]
\centering
\renewcommand{\tabcolsep}{3pt}
{\small
	\caption{Spearman's rank correlation coefficient $\rho$, and Kendall's rank correlation coefficient $\tau$, of occurrence distribution in Figures \ref{histo_occurrence_211}, \ref{histo_occurrence_center} and \ref{histo_occurrence_limb}. Bold values show the deviation from zero in 99\% significance level.}
\begin{tabular}{cccccccccccc}
\hline
& & \multicolumn{2}{c}{All (211 events)} & & \multicolumn{2}{c}{On-disk (63 events)} & & \multicolumn{2}{c}{Limb (148 events)} \\
&  & $\rho$ & $\tau$ &  & $\rho$ & $\tau$ &  & $\rho$ & $\tau$  \\
\cline{2-4} \cline {6-7}\cline{9-10} 
duration -- sigmoid & & {\bf 0.96} & {\bf 0.90}& &  0.85 &0.72 & & 0.79&0.62\\
duration -- CME & & 0.71 & 0.52 && 0.76& 0.65&&0.86&0.71\\
rise time -- sigmoid & & 0.57 & 0.50&& 0.20 & 0.14&&0.35&0.18\\
rise time -- CME & & {\bf 0.98} & {\bf 0.93} && 0.81& 0.69&&{\bf 0.98}&{\bf 0.93}\\
decay time -- sigmoid& & 0.75 & 0.62 && {\bf 0.93} & {\bf 0.82}&&0.44&0.41\\
decay time -- CME & & 0.86 & 0.71 &&0.85& 0.72&&0.86&0.71\\
X-ray flux -- sigmoid &&0.41&0.25 &&0.63& 0.49&&0.26&0.18\\
X-ray flux -- CME&&{\bf 0.98} & {\bf 0.92} &&0.83& 0.68&&{\bf 0.90}&{\bf 0.78}\\
\hline
\label{rho_tau}
\end{tabular}
}
\end{table}

Figure \ref{histo_occurrence_211} shows the sigmoid and CME occurrence as functions of the respective flare properties. The length of the error bar is estimated assuming Poisson noise. Hence, a larger error bar is due to a smaller sample size (e.g., only one event exists at which the rise time ranges between $10^{3.7} < t_{\rm rise} < 10^{3.9}$ s. It can be estimated whether a CME was included. Therefore, the 1-sigma uncertainty of the occurrence in the range equals 1.0). In order to estimate the dependency of the sigmoid and CME occurrence on the respective flare parameter, the histograms were fitted to a linear function by using the least-square method and considering the Poisson error scale. 
Only when the Spearman's rank correlation coefficient $\rho$, and Kendall's rank correlation coefficient $\tau$ are significantly different from 0 in 99\% confidence interval, the fitted lines are shown by the bold line. 
The error value of a slope $R$ is  the 1 $\sigma$ uncertainty.
Regarding sigmoid occurrence, we obtained a slope of $R = 0.43\pm0.22$ for the duration, which is a positive value for a 1$\sigma$ uncertainty. Regarding CME occurrence, we obtained a positive slope for two parameters: $R = 0.40\pm0.12$ for the rise time, and $R = 0.53\pm0.13$ for the soft-X-ray peak flux. 
 
  To compare on-disk and limb events, we derived the sigmoid and CME occurrence for on-disk and limb events as functions of the respective flare properties. We conducted linear fitting, as mentioned above. The results are shown in Figures \ref{histo_occurrence_center} and \ref{histo_occurrence_limb}. Since the total number of events for sigmoids is small (42 events), the occurrence error is large. 
  For the on-disk events, the sigmoid occurrence as a function of the decay time possesses a steep slope ($R = 0.61\pm0.41$). For the limb events, we obtained a positive slope from the CME occurrence as a function of the rise time ($R = 0.45\pm0.14$), and X-ray flux ($R = 0.60\pm0.15$).

\subsection{Linear Correlations Between Parameters}
We investigated the linear correlations between physical parameters by showing the scatter plot. 
Before presenting the result of scatter plot, we show the Pearson's correlation coefficient $C$, Spearman's rank correlation coefficient $\rho$, and Kendall's rank correlation coefficient $\tau$, between two parameters in Table \ref{correlation_rho_tau}. 
Same as Table \ref{rho_tau}, bold values show the deviation from zero in 99\% significance level.
As we can see, there are significant correlation among the duration, decay time, and rise time  for all, on-disk, and limb events. 
Significant correlation is also seen between duration and CME width, decay time and CME width, duration and X-ray flux, and decay time and X-ray flux, respectively for all events.
In the limb events, the correlation between duration and CME width, and decay time and CME width, respectively, does not exist.
 Note that the results of linear fitting are shown later in this section but the linear model may be invalid when the rank correlation coefficient is not significantly different from 0 in 99\% confidence interval. 
\begin{table}[H]
\centering
\renewcommand{\tabcolsep}{3pt}
{\small
\caption{Pearson's correlation coefficient $C$, Spearman's rank correlation coefficient $\rho$, and Kendall's rank correlation coefficient $\tau$, between two parameters. Bold values show the deviation from zero in 99\% significance level.}
\begin{tabular}{cccccccccccccccc}
\hline
 & & \multicolumn{3}{c}{All (211 events)} & & \multicolumn{3}{c}{On-disk (63 events)}
& & \multicolumn{3}{c}{Limb (148 events)} \\
 &  & $C$ & $\rho$ & $\tau$&  & $C$ & $\rho$ & $\tau$ &  & $C$ & $\rho$ & $\tau$  \\
\cline{2-5} \cline{7-9} \cline{11-13}
duration -- rise time & & 0.62 & {\bf 0.63} & {\bf 0.47}& & 0.61 & {\bf 0.54} &{\bf 0.41} & &0.63 & {\bf 0.67}&{\bf 0.51}\\
duration -- decay time & & 0.97 & {\bf 0.96} & {\bf 0.85}& &0.97 & {\bf 0.96} &{\bf 0.87} & & 0.97 & {\bf 0.96}&{\bf 0.84}\\
rise time -- decay time & & 0.45 & {\bf 0.46} & {\bf 0.32}& & 0.45 & {\bf 0.38} &{\bf 0.28} & &0.45 & {\bf 0.49}&{\bf 0.35}\\
duration -- CME speed & & 0.19 & 0.20& 0.15 & & 0.17& 0.22&0.15&&0.19&0.18& 0.14\\
rise time -- CME speed & & 0.16 &0.14& 0.10& &0.18& 0.15&0.09&&0.16&0.13&0.09\\
decay time -- CME speed & & 0.16 & 0.19& 0.13 & &0.13& 0.20&0.12&&0.17&0.18&0.12\\
duration -- CME width &&0.38&{\bf 0.37}&{\bf 0.26}&&0.55&0.42&0.35&&0.30&0.31&0.22\\
rise time -- CME width &&0.27&0.27&0.21&&0.09&0.07&0.07&&0.30&0.31&0.24\\
decay time -- CME width &&0.40&{\bf 0.37}&{\bf 0.27}&&0.60& 0.51&0.42&&0.31&0.30&0.23\\
duration -- X-ray flux &&0.44&{\bf 0.36}&{\bf 0.25}&&0.27&0.22&0.16&&0.51&{\bf 0.42}&{\bf 0.29}\\
rise time -- X-ray flux &&0.14&0.10&0.07&&0.05&-0.04&-0.03&&0.17&0.16&0.11\\
decay time -- X-ray flux &&0.47&{\bf 0.39}&{\bf 0.28}&&0.30&0.24&0.17&&0.54&{\bf 0.47}&{\bf 0.33}\\
\hline
\label{correlation_rho_tau}
\end{tabular}
}
\end{table}

\begin{figure}[H]
\centering
 \includegraphics[width=\columnwidth]{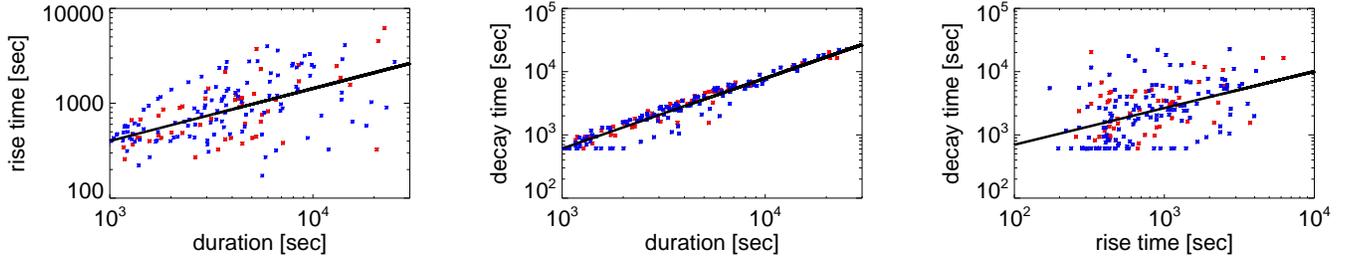}
 \caption{Scatter plot of duration and rise time (left panel), duration and decay time (middle panel), rise time and decay time (right panel). The red and blue points mark on-disk and limb events, respectively.}
 \label{scat_duration}
\end{figure}

\begin{figure}[H]
\centering
 \includegraphics[width=\columnwidth]{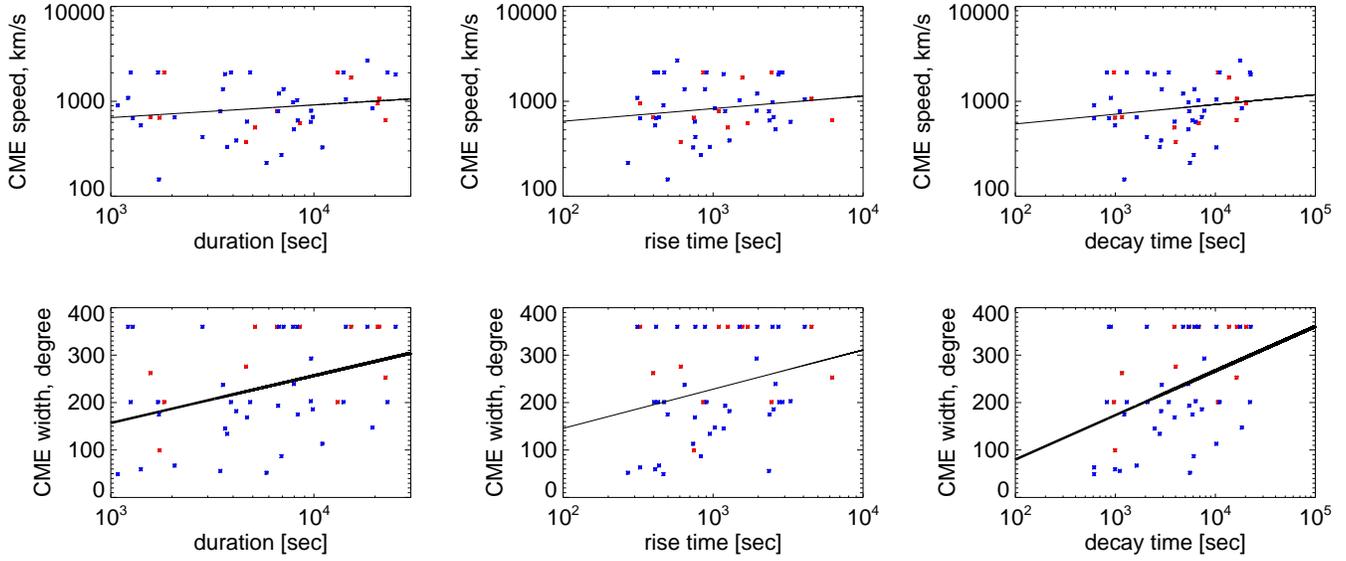}
 \caption{Scatter plot of duration and CME speed (top left), rise time and CME speed (top middle), decay time and CME speed (top right), duration and CME width (bottom left), rise time and CME width (bottom middle), and decay time and CME width (bottom right). The red and blue points mark the on-disk and limb events, respectively.}
 \label{scat_duration_cme}
\end{figure}

\begin{figure}[H]
\centering
 \includegraphics[width=\columnwidth]{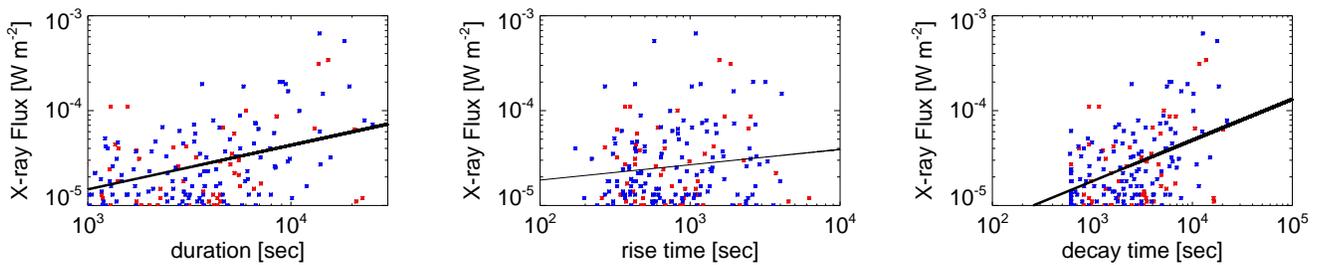}
 \caption{Scatter plot of duration and X-ray class (left panel), rise time and X-ray class (middle panel), and decay time and X-ray class (right panel). The red and blue points mark the on-disk and limb events, respectively.}
 \label{scat_duration_xray}
\end{figure}

Figure \ref{scat_duration} shows the scatter plots of the duration and rise time (left panel), duration and decay time (middle panel), and rise time and decay time (right panel). The red and blue points mark on-disk and limb events, respectively. The correlation coefficients $C$ are shown for all (black), on-disk (red), and limb (blue) events.    In Figures \ref{scat_duration}, \ref{scat_duration_cme}, and \ref{scat_duration_xray}, we show the result of linear fitting, in which bold line is used, when the the values of  $\rho$ and $\tau$ are significantly different from 0 in 99\% confidence interval.
The correlation coefficients of the duration and decay time (middle panel) is extremely large, $C=0.97$. However, $C$ is relatively small ($\sim 0.45$) for the decay and rise time.

Figure \ref{scat_duration_cme} shows the scatter plot of the duration and CME speed (top left), rise time and CME speed (top middle), decay time and CME speed (top right), duration and CME width (bottom left), rise time and CME width (bottom middle), and decay time and CME width (bottom right). 
The correlation coefficients are small and similar for all relations for all events, $C=0.16-0.40$.
In the on-disk events, however, the correlation coefficient between the duration and CME width, and decay time and CME width shows relatively high, $C=0.55, 0.60$. 
The low correlation in the limb events may come from the upper limit of the CME width (360 degree).
There are more events in the limb events than the on-disk events.
As the number of CME events with 360 degree width increases, the correlation coefficient becomes low.

Figure \ref{scat_duration_xray} presents the scatter plot of the duration and X-ray class (left panel), rise time and X-ray class (middle panel), and decay time and X-ray class (right panel).  
The correlation coefficients for all events between the duration and X-ray class, and the decay time and X-ray class, are relatively large: $C=0.44, 0.47$, respectively. 
In contrast, that of rise time and X-ray class is quite small, $C=0.14$. 

\subsection{Association Between Parameters}
Table \ref{ind1} summarizes the occurrence probabilities.
The occurrence probabilities of LXC events and LDEs have roughly half. 
As described above, in 5 of 211 flare events, data is lacking in {\it GOES}.
In addition, certain events occur at decay times of other events, which makes it difficult to evaluate the duration. 
We excluded these events from our analysis.
Thus, 186 events exist in the second row in Table \ref{ind1} instead of 211.
The SGMs are accompanied by flares in approximately half of the events.
The CME occurrence probability is $32.2 \%$, which is slightly lower than that for LXC, LDE, and SGM. 
All occurrence probabilities do not show significant variations among all, on-disk, and limb events.

The contingency tables of each two sets regarding LXC, LDE, SGM, and CME are summarized in Table \ref{ind2}.
The forth and seventh columns show the intersection probabilities and those of negation on both events.
The fifth and sixth columns present the results for the intersection probability of negation on one event.
In the SGM--CME analysis, many unaccompanied SGMs exist (39.2$\%$), but the quantity of CMEs without SGMs is much lower (12.8$\%$).
Many LXC events exist without CMEs (25.4$\%$), but the quantity of CMEs without LXC events is much lower (9.6$\%$).
Further, LDEs without CME are observed ($26.1\%$), but the occurrence probability of CMEs without LDEs is much lower (8.9$\%$).
Nevertheless, no dependency on the observation place could be observed (on-disk or limb) for LXC--LDE, LXC--SGM, and LDE--SGM.
The phi coefficient is shown in the eighth column.                                                                   
We employ 0.2 as a threshold value for a weak association.                                                             
The phi coefficient larger than 0.2 is shown in the bold face in Table \ref{ind2}.
The dependency on the observation place was seen.
In all and limb cases, the phi coefficients of all pairs from LXC--LDE--CME show relatively high values.
On the other hand, those from LDE--SGM--CME are larger than 0.2 but those of LXC--CME and LXC--SGM are not in the on-disk case.

The results of association analysis for the intersection sets are summarized in Table \ref{ind3}. The parameter $P_{\rm ind}$ represents the intersection probability assuming that two sets are independent, $P_{\rm ind}=P(A)\times P(B)$;  $P_{\rm obs}$ corresponds to the intersection of the events according to observations, $P(A \cap B)$ in Table \ref{ind2}; $\Delta P$ is the difference between $P_{\rm ind}$ and $P_{\rm obs}$. The error value $\sigma_{\Delta}$ is obtained from the propagation of the error values assuming independent $\sigma_{\rm ind}$ and $\sigma_{\rm obs}$, i.e., $\sqrt{\sigma_{\rm ind}^2+\sigma_{\rm obs}^2}$.
For all events, LXC--CME and CME--LDE show significant probabilities (1.25 and 1.42), while LXC--SGM and SGM--CME exhibit smaller probabilities (-0.36 and -0.33) compared to other events. 
Although there are no significant probabilities in the on-disk events, LXC-SGM, SGM-CME, and CME-LDE show comparatively higher probabilities.
For limb events, the CME--LXC and CME-LDE values show significant dependency.

\renewcommand{\arraystretch}{0.8}
\begin{table}
\caption{Occurrence Probability of Events}
\begin{tabular}{c@{\hspace{0.3cm}}c@{\hspace{0.8cm}}cccc@{\hspace{0.5cm}}cccc@{\hspace{0.5cm}}cccc@{\hspace{0.5cm}}cccc@{\hspace{0.3cm}}} \toprule
&     & & \multicolumn{2}{c}{All}      & & & \multicolumn{2}{c}{On-Disk} & & & \multicolumn{2}{c}{Limb}    & & \\ 
\cline{4-5} \cline{8-9} \cline{12-13}
&     & & Number    & Probability      & & & Number  & Probability       & & & Number   & Probability      & & \\ 
\midrule
& LXC & & 105 / 211 & 49.8$\pm$4.9$\%$ & & & 32 / 63 & 50.8$\pm$9.0$\%$  & & & 73 / 148 & 49.3$\pm$5.8$\%$ & & \\
& LDE & & 89 / 186 & 47.8$\pm$5.1$\%$ & & & 29 / 53 & 54.7$\pm$10.2$\%$  & & & 60 / 133 & 45.1$\pm$5.8$\%$ & & \\
& SGM & & 58 / 120  & 48.3$\pm$6.3$\%$ & & & 21 / 42 & 50.0$\pm$10.9$\%$ & & & 37 / 78  & 47.4$\pm$7.8$\%$ & & \\
& CME & & 57 / 177  & 32.2$\pm$4.3$\%$ & & & 15 / 51 & 29.4$\pm$7.6$\%$  & & & 42 / 126 & 33.3$\pm$5.1$\%$ & & \\ \bottomrule
\end{tabular}
\label{ind1}
\end{table}
\renewcommand{\arraystretch}{1.0}

\renewcommand{\arraystretch}{0.8}
\begin{table}
\caption{ {\bf Contingency Table of Events and Mean Square Contingency Coefficient}}
\begin{tabular}{ccc@{\hspace{0.4cm}}c@{\hspace{0.6cm}}c@{\hspace{0.6cm}}c@{\hspace{0.6cm}}c@{\hspace{0.6cm}}c@{\hspace{0.4cm}}c} \toprule
        & A    & B    & $N_{\rm tot}$ & ${P(\rm A \cap B)}$ & ${P(\rm \overline{A} \cap B)}$ & ${P(\rm A \cap \overline{B})}$ & ${P(\rm \overline{A} \cap \overline{B})}$ & $\phi$ \\ 
\midrule
All     &      &      &               & & & & & \\
        & LXC  & LDE  & 186           & 28.5$\pm$3.8$\%$ (53) & 19.4$\pm$3.2$\%$ (36) & 18.4$\pm$3.1$\%$ (34) & 33.9$\pm$4.3$\%$ (63) & {\bf 0.25} \\
        & LXC  & SGM  & 120           & 21.7$\pm$4.2$\%$ (26) & 26.7$\pm$4.7$\%$ (32) & 26.7$\pm$4.7$\%$ (32) & 25.0$\pm$4.6$\%$ (30) & 0.07 \\
        & LXC  & CME  & 177           & 22.6$\pm$3.6$\%$ (40) &  9.6$\pm$2.3$\%$ (17) & 25.4$\pm$3.8$\%$ (45) & 42.1$\pm$4.9$\%$ (75) & {\bf 0.31} \\
        & LDE  & SGM  & 108           & 27.8$\pm$5.1$\%$ (30) & 21.3$\pm$4.4$\%$ (23) & 23.2$\pm$4.6$\%$ (25) & 27.8$\pm$5.1$\%$ (30) & 0.11 \\
        & SGM  & CME  & 102           & 13.7$\pm$3.7$\%$ (14) & 12.8$\pm$3.5$\%$ (13) & 39.2$\pm$6.2$\%$ (40) & 34.3$\pm$5.8$\%$ (35) & 0.01 \\
        & CME  & LDE  & 157           & 24.8$\pm$4.0$\%$ (39) & 26.1$\pm$4.1$\%$ (41) &  8.9$\pm$2.4$\%$ (14) & 40.1$\pm$5.1$\%$ (63) & {\bf 0.32} \\ \midrule
On-Disk &   &   & & & & & & \\
        & LXC  & LDE  & 53            & 26.4$\pm$7.1$\%$ (14) & 28.3$\pm$7.3$\%$ (15) & 22.6$\pm$6.5$\%$ (12) & 22.6$\pm$6.5$\%$ (12) & 0.02 \\
        & LXC  & SGM  & 42            & 28.6$\pm$9.0$\%$ (12) & 21.4$\pm$7.1$\%$ (9)  & 21.4$\pm$7.1$\%$ (9)  & 28.6$\pm$8.3$\%$ (12) & 0.14 \\
        & LXC  & CME  & 51            & 15.7$\pm$5.5$\%$ (8)  & 13.7$\pm$5.2$\%$ (7)  & 27.5$\pm$7.3$\%$ (14) & 43.1$\pm$9.2$\%$ (22) & 0.13 \\
        & LDE  & SGM  & 39            & 30.8$\pm$8.7$\%$ (12) & 18.0$\pm$6.8$\%$ (7)  & 18.0$\pm$6.8$\%$ (7)  & 33.3$\pm$9.3$\%$ (13) & {\bf 0.28} \\
        & SGM  & CME  & 34            & 20.6$\pm$7.8$\%$ (7)  &  8.8$\pm$5.1$\%$ (3)  & 32.4$\pm$9.8$\%$ (11) & 38.3$\pm$10.6$\%$ (13) & {\bf 0.22} \\
        & CME  & LDE & 44             & 25.0$\pm$7.5$\%$ (11) & 36.4$\pm$9.1$\%$ (16) &  6.8$\pm$3.9$\%$ (3)  & 31.8$\pm$8.5$\%$ (14) & {\bf 0.24} \\ \midrule
Limb    &   &   & & & & & & \\
        & LXC  & LDE  & 133           & 29.3$\pm$4.7$\%$ (39) & 15.8$\pm$3.5$\%$ (21) & 16.5$\pm$3.5$\%$ (22) & 38.4$\pm$5.4$\%$ (51) & {\bf 0.34} \\
        & LXC  & SGM  & 78            & 18.0$\pm$4.8$\%$ (14) & 29.5$\pm$6.1$\%$ (23) & 29.5$\pm$6.1$\%$ (23) & 23.1$\pm$5.4$\%$ (18) & 0.18 \\
        & LXC  & CME  & 126           & 25.4$\pm$4.5$\%$ (32) &  7.9$\pm$2.5$\%$ (10) & 24.6$\pm$4.4$\%$ (31) & 42.1$\pm$5.8$\%$ (53) & {\bf 0.37} \\
        & LDE  & SGM  & 69            & 26.1$\pm$6.2$\%$ (18) & 23.2$\pm$5.8$\%$ (16) & 26.1$\pm$6.2$\%$ (18) & 24.6$\pm$6.0$\%$ (17) & 0.02 \\
        & SGM  & CME  & 68            & 10.3$\pm$3.9$\%$ (7)  & 14.7$\pm$4.7$\%$ (10) & 41.2$\pm$7.8$\%$ (28) & 33.8$\pm$7.1$\%$ (23) & 0.12 \\
        & CME  & LDE  & 113           & 24.8$\pm$4.7$\%$ (27) & 22.1$\pm$4.4$\%$ (25) & 9.7$\pm$2.9$\%$ (11) & 43.4$\pm$6.2$\%$ (49) & {\bf 0.35} \\
\bottomrule
\end{tabular}
\label{ind2}
\end{table}
\renewcommand{\arraystretch}{1.0}

\renewcommand{\arraystretch}{0.8}
\begin{table}
\caption{\bf Association Analysis for Intersection Sets}
\begin{tabular}{cc@{\hspace{1.5cm}}c@{\hspace{0.8cm}}c@{\hspace{0.8cm}}c@{\hspace{0.8cm}}c} \toprule
    &  & $P_{\rm ind}\pm\sigma_{\rm ind}$ & $P_{\rm obs}\pm\sigma_{\rm obs}$ & $\Delta P \pm \sigma_{\Delta}$ & $\Delta P / \sigma_{\Delta}$ \\ \midrule
All &         &                &                &                &             \\
    & LXC-LDE & $23.8\pm4.9\%$ & $28.5\pm3.9\%$ & $4.7\pm6.3\%$  & 0.75        \\
    & LXC-SGM & $24.2\pm5.5\%$ & $21.7\pm4.2\%$ & $-2.5\pm6.9\%$ & -0.36       \\
    & LXC-CME & $16.1\pm3.7\%$ & $22.6\pm3.6\%$ & $6.5\pm5.2\%$  & $\bold{1.25}$ \\
    & LDE-SGM & $23.1\pm5.5\%$ & $27.8\pm5.1\%$ & $4.7\pm7.5\%$  & 0.63        \\
    & SGM-CME & $15.5\pm4.1\%$ & $13.7\pm3.7\%$ & $-1.8\pm5.5\%$ & -0.33       \\
    & CME-LDE & $15.4\pm3.7\%$ & $24.8\pm4.0\%$ & $9.4\pm5.4\%$  & $\bold{1.74}$ \\ \midrule
On-Disk &     &                &                &                &             \\
    & LXC-LDE & $27.8\pm10.1\%$ & $26.4\pm7.1\%$ & $-1.4\pm12.3\%$ & -0.11        \\
    & LXC-SGM & $25.4\pm10.0\%$ & $34.9\pm9.0\%$ & $9.5\pm13.5\%$ & $0.70$       \\
    & LXC-CME & $14.9\pm6.5\%$ & $15.7\pm5.5\%$ & $-0.8\pm8.5\%$ & -0.09 \\
    & LDE-SGM & $27.4\pm11.1\%$ & $30.8\pm8.9\%$ & $3.4\pm14.2\%$ & 0.24        \\
    & SGM-CME & $14.7\pm7.2\%$  & $20.6\pm7.8\%$ & $5.9\pm10.6\%$ & $0.56$       \\
    & CME-LDE & $16.1\pm7.2\%$ & $25.0\pm7.5\%$ & $8.9\pm10.4\%$ & 0.86 \\ \midrule
Limb&         &                &                &                &             \\          
    & LXC-LDE & $22.2\pm5.5\%$ & $29.3\pm4.7\%$ & $7.1\pm7.2\%$ & 0.99        \\
    & LXC-SGM & $23.4\pm4.4\%$ & $18.0\pm4.8\%$ & $-5.4\pm6.5\%$ & -0.83       \\
    & LXC-CME & $16.4\pm6.6\%$ & $25.4\pm4.5\%$ & $9.0\pm8.0\%$ & $\bold{1.13}$ \\
    & LDE-SGM & $21.4\pm6.3\%$ & $26.1\pm6.2\%$ & $4.7\pm8.8\%$ & 0.53        \\
    & SGM-CME & $15.8\pm5.0\%$ & $10.3\pm3.9\%$ & $-5.5\pm6.3\%$ & -0.87       \\
    & CME-LDE & $15.0\pm4.2\%$ & $24.8\pm4.7\%$ & $9.8\pm6.3\%$ & $\bold{1.56}$ \\ \bottomrule
\end{tabular}
\label{ind3}
\end{table}

\subsection{KS and AD Test} 
We conducted the KS and AD tests and judged the dependency of the sigmoid and CME on the flare parameters. 
We selected 58 events in which a sigmoid existed, and 62 events without sigmoids (see Table \ref{n_event_sgmcme}). 
Moreover, we selected 57 events in which a CME occurred, and 120 events without CMEs (see Table \ref{n_event_sgmcme}).

\begin{figure}[h]
\centering
 \includegraphics[clip,width=0.9\columnwidth]{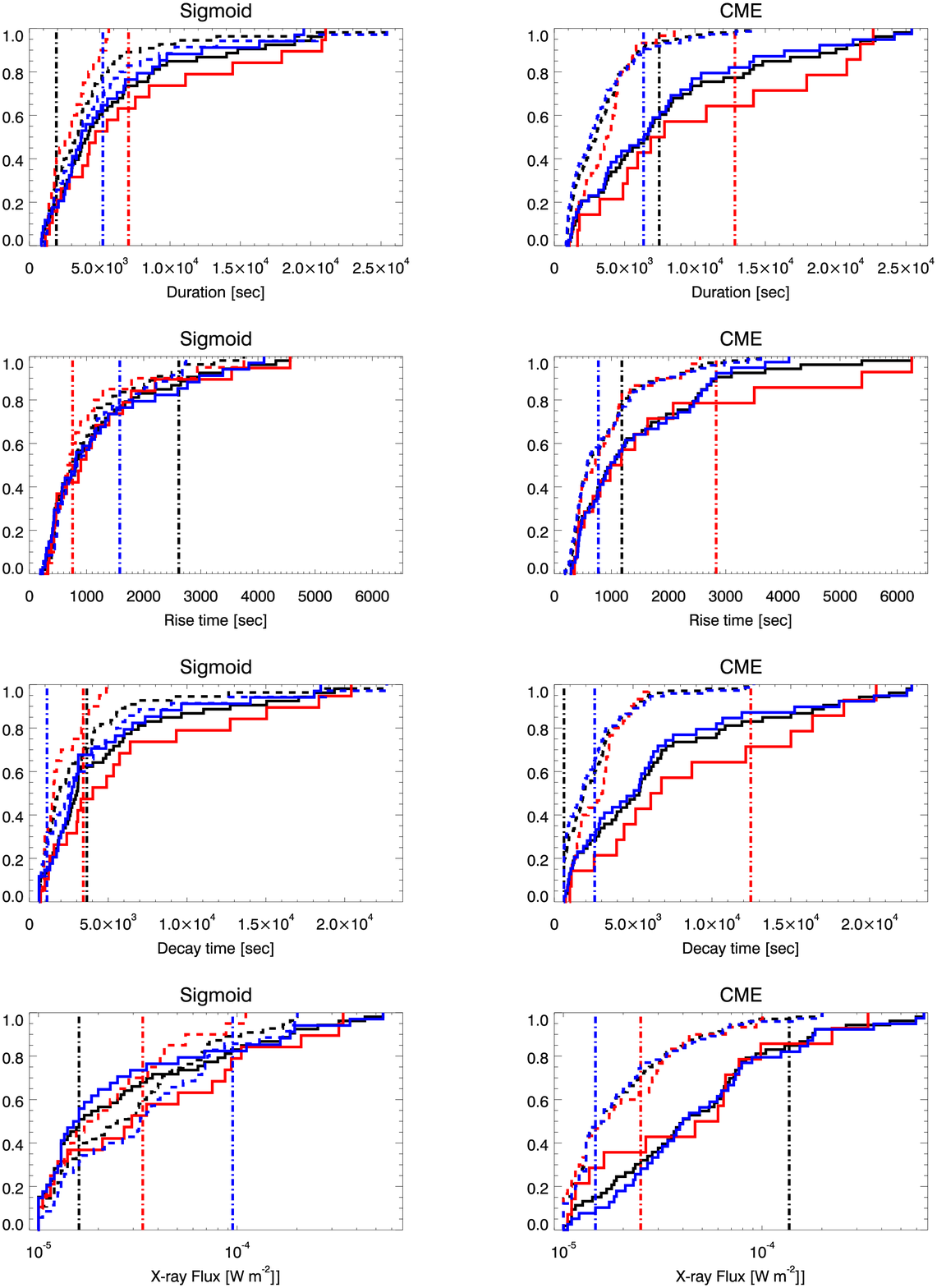}
 \caption{Cumulative distribution functions of sigmoid (left side) and CME (right side) for flare duration, rise time, decay time, and soft-X-ray peak flux. The solid curves represent the CDF for events with a sigmoid (CME), whereas the dashed curves show the CDF for the events without sigmoids (CMEs). The vertical dash--dot lines indicate the maximal deviation between two CDFs. The black, red, and blue lines represent all events, on-disk events, and limb events, respectively.}
 \label{kscdf}
\end{figure}

Figure \ref{kscdf} shows the CDFs. The left and right panels correspond to the sigmoid and CME CDFs, and the black, red, and blue lines represent all events, on-disk events, and limb events, respectively. The solid curves mark the CDF for events in which a sigmoid (CME) occurred, whereas the dashed curves mark the CDF for events without sigmoids (CMEs). The vertical dash--dot lines indicate the maximal deviation between two CDFs. 

 Table \ref{ksvalue_sgm} and \ref{ksvalue_cme} show the values of the KS and AD statistics regarding sigmoid and CME, respectively.  Bold value shows more than the 99\% confidence level ($\alpha < 0.01$).
  According to Table \ref{ksvalue_sgm}, the sigmoid occurrence is not  significantly related to all parameters.
  However, the statistics of both KS and AD tests show relatively larger value in the decay time and duration for all and for the on-disk events, and is less related to the rise time and soft-X-ray peak flux for all and for the on-disk events. 
 Because the number of the sample of the limb events is larger than that of the on-disk events, the value $\sqrt{(n + m)/(nm)}$ of the limb events is lower compared to on-disk events. Therefore, the probability tends to smaller in the limb events than the on-disk events in the KS test.

 According to Table \ref{ksvalue_cme},  in a confidence interval of 99$\%$, the CME occurrence is related to the flare duration and decay time for all, on-disk and limb events and to the soft-X-ray peak flux for all and the limb events. 
 Only for all events, the rise time is related to the CME occurrence in the AD test.
   
\begin{table}[H]
\centering
\renewcommand{\tabcolsep}{3pt}
{\small
\caption{KS and AD statistic for each condition regarding SGMs. Bold value shows more than the 99\% confidence.}
\begin{tabular}{cccccccccc}
\hline
 & & \multicolumn{2}{c}{All (211 events)} & & \multicolumn{2}{c}{On-disk (63 events)}
& & \multicolumn{2}{c}{Limb (148 events)} \\
 & & KS statistic & AD statistic & & KS statistic & AD statistic & & KS statistic & AD statistic \\
\cline{3-4} \cline{6-7} \cline{9-10}
duration & & 0.20 &1.49  & & 0.43 &3.2  & & 0.12 & -0.78 \\
rise time & & 0.12 & -0.18 & & 0.23 & -0.58 & & 0.18 &  -0.23 \\
decay time & & 0.26 & 2.2 & & 0.43 &  3.7& & 0.17 &  -0.60\\
X-ray flux & & 0.13 &  -0.27& & 0.28 & 0.24 & & 0.31 &  1.2\\
\hline
\label{ksvalue_sgm}
\end{tabular}
}
\end{table}

\begin{table}[H]
\centering
\renewcommand{\tabcolsep}{3pt}
{\small
\caption{KS and AD statistic for each condition regarding CMEs.Bold value shows more than 99\% confidence.}
\begin{tabular}{cccccccccc}
\hline
 & & \multicolumn{2}{c}{All (211 events)} & & \multicolumn{2}{c}{On-disk (63 events)}
& & \multicolumn{2}{c}{Limb (148 events)} \\
 & & KS statistic & AD statistic & & KS statistic & AD statistic & & KS statistic & AD statistic \\
\cline{3-4} \cline{6-7} \cline{9-10}
duration & & {\bf 0.44} & {\bf 18.0} & & {\bf 0.59} & {\bf 7.0} & & {\bf 0.42} & {\bf 11.6}  \\
rise time & & 0.23 &{\bf 5.0}  & & 0.30 & 0.38 & & 0.26 &3.7  \\
decay time & & {\bf 0.43} &{\bf 18.3}  & & {\bf 0.59} & {\bf 7.1} & & {\bf 0.40} & {\bf 12.0}\\
X-ray flux & & {\bf 0.41} & {\bf 17.1} & & 0.43 & 2.3& & {\bf 0.49} &  {\bf 15.0}\\
\hline
\label{ksvalue_cme}
\end{tabular}
}

\end{table}



\section{Discussion}
\subsection{CME}

The CME occurrences as functions of duration, rise time, and decay time exhibit positive slope in Figure \ref{histo_occurrence_211}, while the duration and the decay time did not show significant dependence in the 99 $\%$ confidence level.
\cite{2017ApJ...834...56T} also compared the histograms of the duration and decay time for eruptive and noneruptive events. In their results, the decay time histogram for eruptive events shows large values for the long decay time. 
On the other hand the histograms of duration of \cite{2017ApJ...834...56T} shows little difference between eruptive and noneruptive histograms, whereas the CME occurrence as function of the duration shows a positive slope in our study.
Their conclusion that a longer decay time tends to produce CMEs is consistent with our result. 
However, the decay time and the duration show the similar trend in our study.
 This is due to the different definition of decay time. 
 They define the decay time as the $e$-folding time. Therefore, their duration is more prone to being affected by the rise time than the decay time obtained using our definition. In our definition, the rise time is an order of magnitude shorter than the decay time. The duration is determined largely by the decay time from eqn. (3). The relation between CME occurrence and decay time can be understood by examining the relation between decay time and flaring loop length. It is known that the decay time depends linearly on the flaring loop length (regarding the global thermodynamic time scale) \citep{2002ASPC..277..103R}:
\begin{equation}
t_{\rm dec}\sim \frac{120 L_{9}}{\sqrt{T_{7}}} ,
\label{reale}
\end{equation}
where $L_9$ and $T_7$ are loop length and temperature in $10^9$ cm and $10^7$ K, respectively. When a CME occurs, the reconnected field lines, i.e., flare loops, must be large because the reconnection point becomes higher and higher as the flare proceeds. Consequently, the decay time tends to be large. 

In limb events, the CME occurrence shows a strong dependency on the X-ray flux ($R = 0.60 \pm 0.15$), which is consistent with the results in \cite{2006ApJ...650L.143Y}. They showed that the value of the slope is $R = 0.332$ for X-ray flux without on-disk events. 
The value of their study is different in 1$\sigma$ uncertainty of our results. 
We suspect that this is mainly due to different event numbers. 
There exist 16 X-class flares in our study, whereas 98 X-class flares exist in \cite{2006ApJ...650L.143Y}. 
Our results indicate that CMEs are associated with approximately all events whose X-ray flux is larger than $10^{-3.9} \ \rm{W} \rm{m}^{-2}$. This is also consistent with the results in \cite{2006ApJ...650L.143Y}.  

In Figure \ref{scat_duration_cme}, less correlations exist between duration, rise time, CME width, and CME speed. The correlation of CME speed and duration measures 0.19, although \cite{2017ApJ...834...56T}  reported a correlation coefficient of 0.50. It should be noted that they selected events at the heliographic coordinates less than $45^{\circ}$ from the disk center, which makes the CME speed uncertain. Furthermore, more CME events (57 events) exist in our study than in theirs (32 events). Therefore, our results indicating a lower correlation between duration and CME speed should be more reliable. In addition, there exists more sophisticated way deriving the CME speed by \cite{2010ApJ...715..493L}, who studied azimuthal property by using stereoscopic observations. The application of this method to the statistical study will be a future task.
Regarding the CME width, it seems that the correlation coefficients are larger compared to those of the CME speeds. This results from the fact that large CME sizes cause larger flare loops and a longer duration and decay time, as discussed above. 

In Table \ref{ind2}, LXC-CME and LDE-CME  show relatively high association ($\phi >0.2$) in all events and limb events, while LXC-CME does not show high association in on-disk event.
 The less association between LXC and CME in on-disk events might be due to small number of events.
This result indicates that the relation between LDE and CME is stronger than that between LXC and CME.
This might be also affected by the detection uncertainty of each phenomenon and structure. 
In other words, limb events have less uncertainty for CME detection, while it is difficult to associate CMEs with flares for on-disk events. 
Our association analysis for intersection sets in Table \ref{ind3} indicates a high dependency for LXC--CME ($\Delta P/\sigma_\Delta=1.25$) and CME--LDE ($\Delta P/\sigma_\Delta=1.74$) for all events. This is consistent with the results of the occurrence rate in Figure \ref{histo_occurrence_211}. 
However, several relations show a dependency on the position of the flare event. 

The KS and AD tests indicate that in a confidence interval of 99$\%$, the CME is related to longer duration and decay time. 
Further, for the limb and all events, the X-ray flux is significantly related to the CMEs. 
Thus, the relation between CME association and X-ray flux strengthens in limb events, which reflects that CMEs are more easily detected at the solar limb than inside the solar disk.

\subsection{Sigmoids}
In Figure \ref{histo_occurrence_211}, as in the case of SGM occurrence, the positive values of the slope $R = 0.43 \pm 0.22$ for the duration. Thus, the duration tends to increase when a sigmoidal structure exists. A sigmoidal structure means highly twisted magnetic field lines and is usually formed in the large active regions. Hence, sigmoidal structures tend to have long field lines and can lead to long-duration flares from the discussion above. 
From Figures \ref{histo_occurrence_center} and \ref{histo_occurrence_limb}, the sigmoid occurrence as a function of X-ray flux seems to exhibit a positive slope, although their dependencies do not show significant correlations.
 A sigmoidal structure implies a large free energy in the 3D magnetic field. 
 Naturally, a large free energy can lead to large flares. 
 Our results, however, do not  show the significant evidence that a larger free energy is needed to produce larger flares for on-disk events. 


From KS and AD tests, we did not find the relation between flare parameters and sigmoid existence in 99\% confidence interval. However, there were weak relations between some parameters. For all events (black line), the sigmoid existence is related to a longer decay time (the probability is 0.045) and for the on-disk events (red line), the sigmoid existence is related to a longer duration and longer decay time (the probabilities are 0.037 and 0.035). In contrast, for limb events (blue line), the sigmoid existence is not related to the duration, rise time and decay time due to the sigmoid detection uncertainty regarding active regions far away from the disk center. Moreover, according to the CDF curves in the panel of the fourth row on the left side of Figure \ref{kscdf}, if a sigmoid exists, the X-ray flux becomes large only for on-disk events (red line).

In this study, we judged sigmoid existence by our eyes with a help of the method of the detection of the bright region described in Sec. 3.1.3. We plan to develop a fully automatic sigmoid detection method by applying automatic thinning processing to the sigmoidal region. 
Although the thinning processing of the spread area is difficult in general,  one of the promising method is OCCULT-2 developed by \cite{2013Entrp..15.3007A}.

\subsection{Relation between CMEs and sigmoids}
The appearance of X-ray sigmoids may have been considered as an indicator for eruptive flares \citep{1999GeoRL..26..627C}.
In their results, 50\% of non-sigmoidal region produced eruptive events, while 84\% of sigmoidal region produce eruptive events.
Our association analysis in Tables \ref{ind2} and \ref{ind3}, however,  indicate that there is less dependency of sigmoids on CMEs in all and the limb events.
As discussed above, we suspect that in the limb events, sigmoidal structures are difficult to identify due to projection effects.
In the on-disk events, on the other hand, the CME-SGM relation shows comparatively high dependency ($\phi >0.2$) in Table \ref{ind2}, while it is not significant in Table \ref{ind3}.
  Our results in table \ref{ind2} show only 39\% of sigmoidal structures produce CMEs in the on-disk events.
These results show that the sigmoidal structure does not necessarily lead to the eruptive flare.
The result in Table \ref{ind2} also supports this statement. Focusing on the on-disk events, while 20.6 $\%$ of the flare events are with both sigmoids and CMEs, 32.4 $\%$ of the flare events have sigmoidal structure but do not produce CMEs.
Another important result is that there is extremely little number of the flare events with CMEs and without sigmoids.
This result shows that the sigmoidal structure is necessary factor (but not sufficient) for the occurrence of CMEs.
 The methods for identifying sigmoidal structures may result in the derivation in our results from those of \cite{1999GeoRL..26..627C}.
For example, smaller sigmoidal structures  shown in the bottom panel of Figure \ref{fig:sgm_ex} were identified in our study.  
This might decrease non-sigmoidal but eruptive events.  

\cite{2004A&A...422..337T} studied the relation between post-eruptive arcade (PEA) and CMEs and found that almost all post-eruptive arcades as observed by SoHO/EIT were associated with CMEs. 
They also studied the length of PEA at each latitude. It would be an interesting future topic to study the relation between the length of sigmoid and PEA.

\section{Summary}
In summary, we performed statistical analyses to compare the dependency among the parameters (sigmoids existence, CMEs existence, duration, decay time, rise time, X-ray peak flux, CME speeds, and CME width) based on  five analyses; occurrence rate analysis, linear-correlation analysis, association analysis, the KS test, and the AD test. 
 Our statistical analysis reproduced some results consistent with previous studies.
  The dependency of the CME occurrence on the X-ray flux is consistent with \cite{2006ApJ...650L.143Y}.
  The result that longer decay time flares tend to produce CMEs is also consistent with \cite{2017ApJ...834...56T} .
  Regarding sigmoidal structures, \cite{1999GeoRL..26..627C} investigated the eruptivity of sigmoids based on soft-X-ray images only. 
 In contrast, our statistical analysis shows a dependency between sigmoids and CMEs by comparing soft X-ray images with a comprehensive CME catalog from {\it SoHO}/LASCO. 
  One of our new results is that the dependency between sigmoids and CMEs  changes with the location of the events, probably due to the uncertainty in the detection of sigmoidal structures. 
  Another new result is that while the flare events with sigmoids do not necessarily produce CMEs,  the CMEs without sigmoids are very few. 

In terms of space weather, the important subject to predict is the CME occurrence. 
We conclude that the sigmoidal structure shows comparatively strong dependency on the CME occurrence in on-disk events, whereas the decay time, duration, and X-ray class depend on the CME occurrence in on-disk and limb events. Since CMEs produced on-disk have a high possibility to have a great impact on the Earth environment, sigmoidal structures are very useful to predict CMEs approaching Earth. In addition, sigmoidal structures are often created more than tens of hours before the flare onset, thereby enabling an early prediction of CMEs. However, regarding limb events, the decay time, duration, and X-ray class help to predict CMEs, which can only be used after the occurrence of solar flares.

{\bf Acknowledgements}
We thank the referee for constructive criticisms and valuable comments, allowing us to improve the manuscript significantly. 
{\it Hinode} is a Japanese mission developed and launched by ISAS/JAXA, in collaboration with NAOJ as domestic partner, and NASA and STFC (UK) as international partners. The mission is operated by these agencies in cooperation with ESA and NSC (Norway).
This study was carried out using the Hinode Flare Catalogue(https://hinode.isee.nagoya-u.ac.jp/flare\_catalogue/), which is maintained by ISAS/JAXA and the Institute for Space--Earth Environmental Research (ISEE) at Nagoya University. A part of this study was carried out by the joint research program of ISEE.
The CME catalog is generated and maintained at the CDAW Data Center at NASA and The Catholic University of America in cooperation with the Naval Research Laboratory (USA). SOHO is an international cooperation project between ESA and NASA. 
This work was supported by MEXT/JSPS KAKENHI Grant Number JP15H05814.
\bibliographystyle{apj}

\end{document}